\documentclass{jfm}

\usepackage{natbib}

\usepackage{graphicx}
\usepackage{amsmath,amssymb}
\usepackage{xcolor}

\newcommand{\Ra}{\mathit{R}}      
\newcommand{\Rar}{\mathit{Ra}}    
\newcommand{\Ek}{\mathit{E}}

\newcommand{\ez}{\boldsymbol{e}_z}
\newcommand{\bu}{\boldsymbol{u}}

\title{Wall modes in rotating penetrative internally heated convection}

\author{Gabriel A. Tarditti\aff{1}, Rodolfo Ostilla-M\'onico\aff{1}  \corresp{\email{rodolfo.ostilla@uca.es}} \and Ali Arslan\aff{2}}

\affiliation{
\aff{1}Dpto.~Ing. Mec\'anica y Dise\~no Industrial, Escuela Superior de Ingenier\'ia, Universidad de C\'adiz, Av.~de la Universidad de C\'adiz 10, 11519 Puerto Real, Espa\~na 
\aff{2} Institute of Geophysics, ETH Zurich, 8092 Zurich, Switzerland
}

\shorttitle{Wall modes in rotating penetrative internally heated convection}
\shortauthor{G. A. Tarditti, R. Ostilla-M\'onico \& A. Arslan}

\begin{document}

\maketitle

\begin{abstract}
Rotating internally heated convection has been studied almost exclusively in laterally periodic domains, which cannot support the sidewall-attached modes that appear for sufficient rotation rates in rotating Rayleigh--B\'enard convection in containers. We conduct direct numerical simulations of internally heated convection with isothermal plates and an adiabatic sidewall in cylinders of aspect ratio $\Gamma = 1/2$, $1$, and $2$ at $\Ra = 10^8$ and $\Pran = 1$, comparing them with a doubly periodic layer. When rotation is strong enough to stabilise the flow in a doubly periodic layer at the same $R$ ($\Ek \lessapprox 2 \times 10^{-5}$), the cylinders continue transporting heat via a wall mode down to $\Ek = 5.62 \times 10^{-6}$. This mode consists of a wave attached to the sidewall, with a radial thickness of roughly $0.03$ layer depths independent of $\Gamma$, leaving a stationary interior. Although unstable stratification is confined to the upper half of the conductive state, the mode also penetrates the lower half while retaining stable mean stratification in the measured lower-layer interval. Its azimuthal wavenumber scales roughly with $\Gamma$, yielding a wall parallel wavelength on the order of the layer depth, and it drifts retrograde as a rigid wave slower in wider cylinders and at lower $\Ek$. Analogous to rotating Rayleigh--B\'enard convection, the wall modes reported here extend the observed range of sustained convection relative to the periodic reference and may complicate experimental comparisons.
\end{abstract}

\begin{keywords}
xxx
\end{keywords}

\section{Introduction}\label{sec:intro}
Rotation alters how a fluid subject to buoyant forces first becomes unstable. For example, in rotating Rayleigh--B\'enard convection (RRBC), the Taylor--Proudman theorem constrains vertical motion and the onset of bulk convection is delayed scaling with rotation rate as $\Rar\sim\Ek^{-4/3}$ \citep{Chandrasekhar1961}. Here the Rayleigh number $\Rar$ measures the ratio of destabilising buoyancy to diffusive effects and the Ekman number $\Ek$ the ratio of viscous to Coriolis effects. A lateral boundary relaxes this constraint locally, so in confined containers such as cylinders the first instability is not a bulk mode but a mode attached to the sidewall, which appears well below the bulk threshold \citep{Ecke1992,Zhong1993,Kuo1993,Goldstein1993}. Asymptotic theory describes this sidewall mode as a travelling wave attached to the wall and spanning the full depth of the fluid, with an onset scaling as $\Ek^{-1}$ for adiabatic walls \citep{Kuo1993,Herrmann1993}. The wall mode persists well above onset and simulations show robust wall states that coexist with the rapidly rotating bulk, in turn affecting global heat transport \citep{Horn2017,Favier2020}. Therefore, wall modes remain a central consideration when interpreting laboratory experiments \citep{Ecke2023}.

Classical results of wall modes are based on a fluid heated at the boundaries, where the horizontally and time-averaged vertical heat flux is constant with height. Rotating internally heated convection (RIHC), relevant to the flows of planetary and stellar interiors \citep{Goluskin2016}, breaks this vertical uniformity. With uniform volumetric heating and two plates held at equal fixed temperatures, the conductive state reaches its temperature maximum at mid-depth, resulting in an unstably stratified upper and stably stratified lower half. A wall mode must then either remain trapped in the unstable region or penetrate into the stable layer, where rotation and stratification support inertial gravity waves. It is therefore unknown whether a wall mode exists in RIHC, and whether it persists at rotation rates where the bulk is stable, as it does in RRBC. Previous analytical and numerical studies of RIHC with cooled plates established the bulk onset scaling ($\Ra\sim\Ek^{-4/3}$), the effects of altering the Prandtl number and the changes in flux asymmetries \citep{Arslan2024,Ostilla2025,Ostilla2026}, but only in horizontally periodic domains, which cannot support sidewall modes. Rotating internally heated flows have also been studied in spherical geometry \citep{Jones2000,Kaplan2017,Seeley2025}, but plate bounded configurations with lateral walls remain unexamined.

Here, we use the doubly periodic layer as a bulk mode benchmark to isolate the specific influence of the lateral wall. We perform direct numerical simulations (DNS) of RIHC in a cylinder with adiabatic sidewalls and isothermal plates at $\Pran=1$ and $\Ra=10^8$, varying the Ekman number for aspect ratios $\Gamma=1/2$, $1$ and $2$, and compare with the periodic layer of \citet{Ostilla2025}, which uses the same numerical method, boundary conditions and Rayleigh number. Specifically, we address three questions: (i) does a sidewall attached mode exist in RIHC, and does it persist at rotation rates where the periodic layer no longer convects; (ii) is the mode confined to the unstably stratified upper half or does it penetrate the stable lower half and (iii) how do the Ekman number and the aspect ratio select the azimuthal wavenumber and precession drift of the mode? Calculating the onset thresholds of the mode through linear stability analysis are left for future work.

\section{Formulation and numerical method}\label{sec:methods}

We consider a Boussinesq fluid of kinematic viscosity $\nu$, thermal diffusivity $\kappa$, thermal expansion coefficient $\alpha$, density $\rho$ and specific heat $c_p$, between two horizontal plates a distance $d$ apart. The layer rotates about the vertical axis at angular velocity $\Omega$, gravity acts along $-\ez$, and the fluid is heated uniformly at a volumetric rate $H$. The Froude number is taken to be negligible, so the centrifugal acceleration is absorbed into the pressure. Lengths are scaled with $d$, time with $d^2/\kappa$, velocity with $\kappa/d$ and temperature with $d^2H/(\kappa\rho c_p)$, following \citet{Ostilla2025}. The non-dimensional equations are
\begin{align}
  \nabla\!\cdot\!\bu &= 0, \label{eq:cont}\\
  \partial_t\bu + \bu\!\cdot\!\nabla\bu + \Pran\,\Ek^{-1}\,\ez\times\bu &= -\nabla p + \Pran\,\nabla^2\bu + \Pran\,\Ra\,T\,\ez, \label{eq:mom}\\
  \partial_t T + \bu\!\cdot\!\nabla T &= \nabla^2 T + 1, \label{eq:temp}
\end{align}
with control parameters
\begin{equation}
  \Ra = \frac{g\alpha H d^5}{\rho c_p\,\nu\,\kappa^2},\qquad
  \Ek = \frac{\nu}{2\Omega d^2},\qquad
  \Pran = \frac{\nu}{\kappa}=1 .
\end{equation}

\noindent The strength of rotation relative to buoyancy is measured by the convective Rossby number $\mathit{Ro}=\Ek\sqrt{\Ra/\Pran}$. 

The Rayleigh number is fixed at $\Ra = 10^8$ throughout, while the Ekman number $\Ek$ and aspect ratio $\Gamma$ are varied. Two geometries are examined. The first is a cylinder of diameter $D$ and aspect ratio $\Gamma = D/d$. No-slip velocity conditions are applied on all rigid boundaries. Both the bottom ($z=0$) and top ($z=1$) plates are isothermal with $T=0$, and the sidewall at $r=\Gamma/2$ is adiabatic ($\partial_r T = 0$). At $\Ra=10^8$, we simulate Ekman numbers in the range $\Ek \in [3.16\times10^{-6},\,3.16\times10^{-2}]$, alongside a non-rotating reference case, for aspect ratios $\Gamma = 1/2$, $1$, and $2$. The second geometry is the Cartesian doubly periodic layer of \citet{Ostilla2025}, bounded vertically by isothermal plates. The non-dimensional horizontal domain size $L_x = L_y = \pi$ corresponds to an effective aspect ratio $\Gamma = \pi$ at which quantities become $\Gamma$ independent; results for this geometry are taken directly from that study.

In both geometries, the conductive state is horizontally uniform and given by
\begin{equation}
  T_c(z) = \tfrac{1}{2}\,z\,(1-z),
\end{equation}
reaching its maximum $T_c = 1/8$ at mid-depth ($z=1/2$). Consequently, the lower half of the layer is stably stratified, whereas only the upper half is unstably stratified. The conductive profile is identical in both configurations, with domain-averaged temperature $\langle \overline{T_c} \rangle = 1/12$, where $\langle\cdot\rangle$ denotes a horizontal-plane and time average and the overbar $\overline{(\cdot)}$ denotes a vertical average. Because $T_c(z)$ is independent of $r$, the adiabatic sidewall does not alter the conductive state, ensuring that any horizontal structure originates entirely from the flow.

The governing equations are integrated using the second-order finite difference solver AFiD \citep{VanDerPoel2015}, validated for internally heated convection in periodic domains by \citet{Ostilla2025}, and its cylindrical coordinate counterpart \citep{Verzicco1996}, extensively validated for cylindrical Rayleigh--B\'enard convection \citep{Stevens2010}. Time integration employs a third-order Runge--Kutta scheme for nonlinear terms and a second-order implicit Crank--Nicolson scheme for linear terms. The coordinate singularity at the cylinder axis $r=0$ is avoided by solving for the radial flux variable $q_r = r u_r$ \citep{Verzicco1996}.

Discretisation in the azimuthal direction is uniform, whereas radial and axial grids are clustered near rigid boundaries. For aspect ratios $\Gamma = 1/2$, $1$, and $2$, the azimuthal and radial grid dimensions $(N_\theta \times N_r)$ are set to $128 \times 64$, $256 \times 64$, and $512 \times 128$, respectively, while $N_z$ ranges between $128$ and $384$ points as $\Ek$ decreases. These grid densities mirror the spatial resolution criteria established in \citet{Ostilla2025}. The grids resolve both the horizontal Ekman layers of thickness $O(\Ek^{1/2})$ and the sidewall boundary layer where the wall mode resides. Numerical accuracy is confirmed using the exact global balances $\overline{\langle|\boldsymbol{\nabla}\mathbf{u}|^2\rangle} = \Ra\,\overline{\langle wT\rangle}$ and $\overline{\langle|\boldsymbol{\nabla} T|^2\rangle} = \overline{\langle T\rangle}$. Each simulation is initialised from a fully developed non-rotating convective state at matching $\Ra$ and $\Gamma$, with rotation introduced impulsively, and are integrated until convective transients decay and statistical stationarity is achieved \citep{Ostilla2025}. Convergence is verified by requiring that the time average of $\langle wT\rangle$ over the second half of the run matches that of the full stationary window to within $1\%$. Because simulations are initialised from finite-amplitude states, they do not pinpoint linear onset thresholds: the minimum $\Ek$ supporting a sustained wall mode may differ from the linear critical threshold if the primary instability is subcritical.

\section{Results}\label{sec:results}

\begin{figure}
  \centering
  \begin{minipage}{0.325\textwidth}
    \centering
    \includegraphics[width=\textwidth]{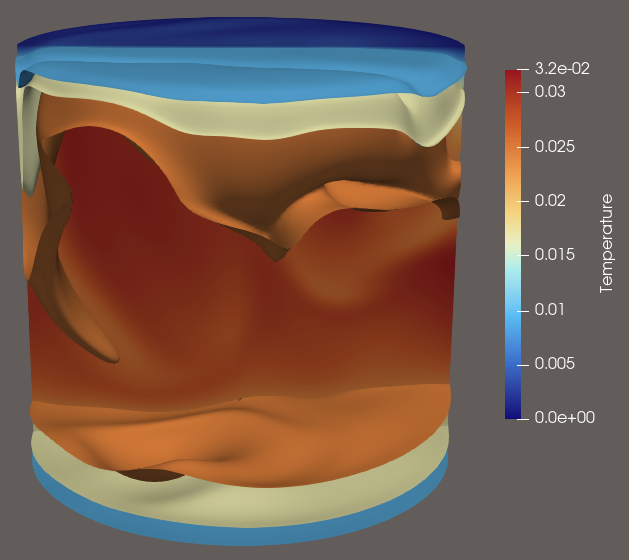}\\
    (a)
  \end{minipage}\hfill
  \begin{minipage}{0.325\textwidth}
    \centering
    \includegraphics[width=\textwidth]{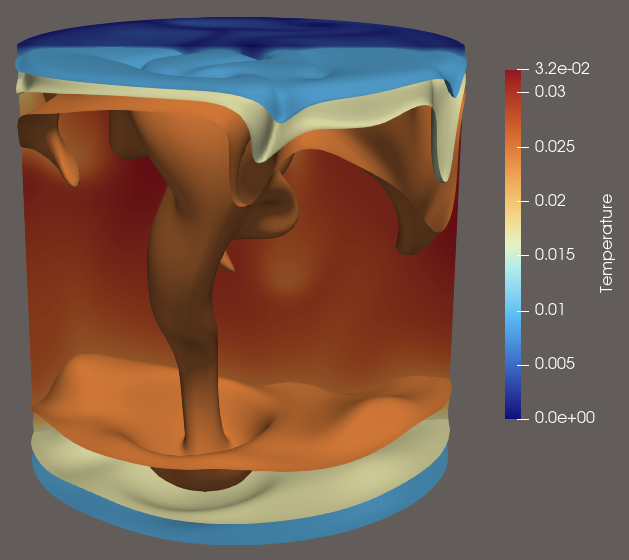}\\
    (b)
  \end{minipage}\hfill
  \begin{minipage}{0.325\textwidth}
    \centering
    \includegraphics[width=\textwidth]{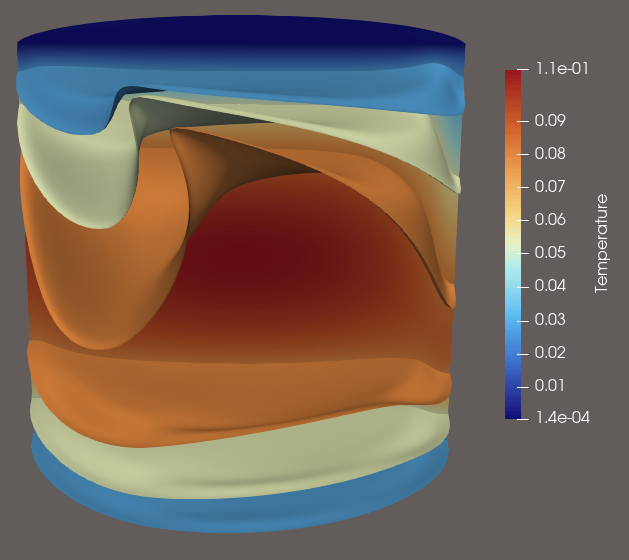}\\
    (c)
  \end{minipage}
\caption{Three-dimensional instantaneous temperature contours $T$ at $\Gamma=1$ and $\Ra=10^8$ for ($a$) the non-rotating case, ($b$) $\Ek=3.16\times10^{-3}$ ($\mathit{Ro}^{-1}=0.032$), where the flow is organised into Taylor columns, and ($c$) $\Ek=10^{-5}$ ($\mathit{Ro}^{-1}=10$), where the interior remains near-conductive and motion is confined to a sidewall mode. Each panel uses an independent color scale.}
  \label{fig:visual}
\end{figure}

Figure~\ref{fig:visual} shows instantaneous temperature fields in the $\Gamma=1$ cylinder for three Ekman numbers. Without rotation (figure~\ref{fig:visual}a), the flow exhibits a turbulent convective structure with plumes emitted from the unstable top which stir the stable bottom. At $\Ek=3.16\times10^{-3}$ (figure~\ref{fig:visual}b), rotation organises the downward plumes into Taylor columns, similar to the behaviour in a periodic layer. At $\Ek=10^{-5}$ (figure~\ref{fig:visual}c), the flow is localised near the sidewall in a wall mode that extends deep enough to penetrate the stably stratified lower layer; the interior is nearly conductive, overall temperatures are higher, and velocity fluctuations are attenuated.

\begin{figure}
  \centering
  \begin{minipage}{0.49\textwidth}
    \centering
    \includegraphics[width=\textwidth]{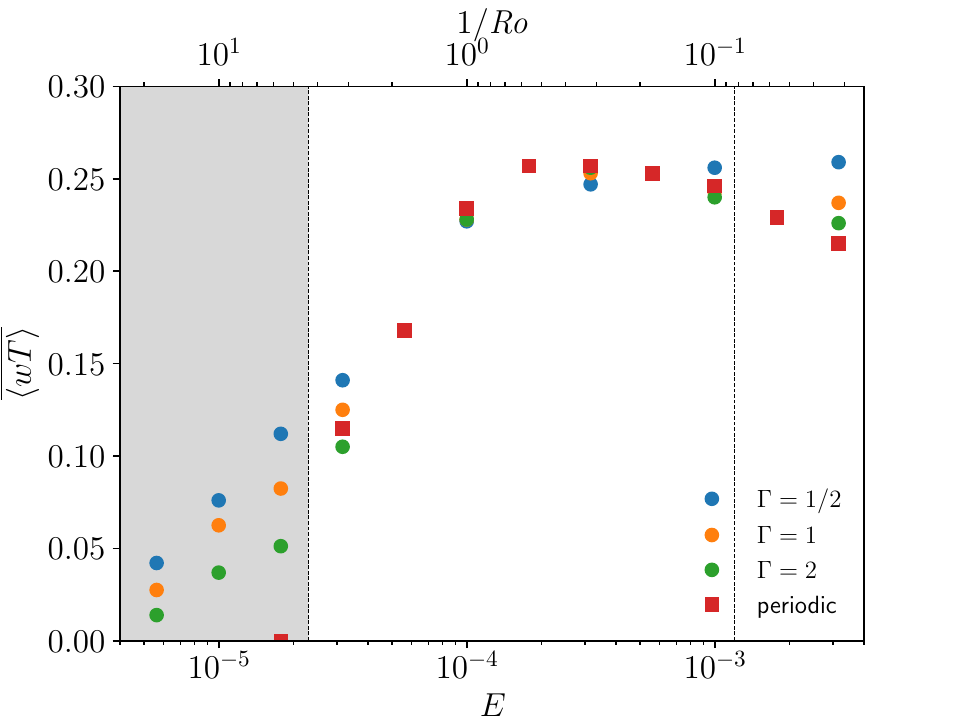}\\
    (a)
  \end{minipage}\hfill
  \begin{minipage}{0.49\textwidth}
    \centering
    \includegraphics[width=\textwidth]{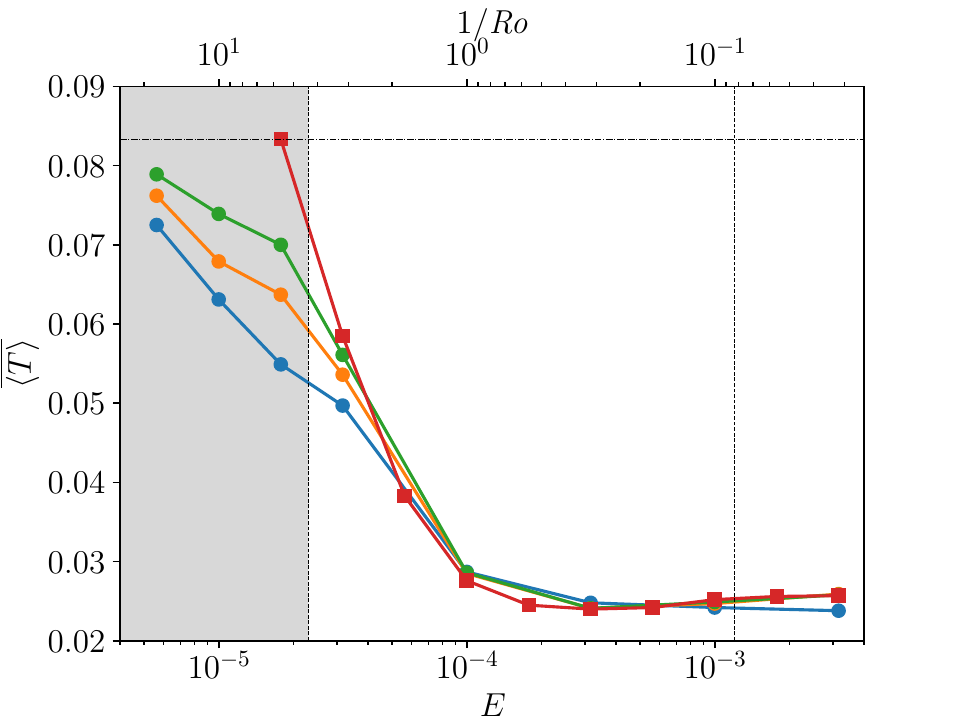}\\
    (b)
  \end{minipage}
  \caption{(a) Convective heat flux $\overline{\langle wT\rangle}$ and (b) mean temperature $\overline{\langle T\rangle}$ as functions of the Ekman number (bottom axis) and of the inverse convective Rossby number $\mathit{Ro}^{-1}$ (top axis) at $\Ra=10^8$, for the cylinders and for the doubly periodic layer \citep{Ostilla2025}. The grey region, $\Ek\lesssim2.3\times10^{-5}$ ($\mathit{Ro}^{-1}\gtrsim4.3$), marks the estimated range in which the periodic layer is conductive and the cylinders sustain a wall mode. The dashed vertical line at $\Ek\approx1.2\times10^{-3}$ ($\mathit{Ro}^{-1}\approx0.083$) marks the transition from rotationally affected (left) to rotationally unaffected (right) flow. The conductive values are $\overline{\langle wT\rangle}=0$ and $\overline{\langle T\rangle}=1/12$ (horizontal dash-dotted line in (b)).}
  \label{fig:flux}
\end{figure}

We first examine global transport properties. Figure~\ref{fig:flux} shows the domain-averaged convective heat flux $\overline{\langle wT\rangle}$ and mean temperature $\overline{\langle T\rangle}$ versus $\Ek$ for the three cylinders and the periodic reference layer. Corresponding conductive baseline values are $0$ and $1/12$, respectively. Because the purely conductive heat flux through each horizontal plate is $1/2$, $\overline{\langle wT\rangle}$ directly determines the fractions of heat removed through the top and bottom boundary plates: $F_T = 1/2 + \overline{\langle wT\rangle}$ and $F_B = 1/2 - \overline{\langle wT\rangle}$, satisfying $F_T + F_B = 1$.

Three dynamical regimes emerge. For weak rotation, $\Ek \gtrsim 1.2\times10^{-3}$ ($\mathit{Ro}^{-1} \lesssim 0.083$), $\overline{\langle wT\rangle}$ and $\overline{\langle T\rangle}$ remain close to their non-rotating values, which show a slight $\Gamma$-dependence: $\overline{\langle wT\rangle} = 0.238$ and $0.227$ for $\Gamma=1$ and $2$, against $0.259$ for the narrow cylinder ($\Gamma=1/2$), likely reflecting the confinement of large scale non-rotating flow structures. As rotation increases, the system enters the rotationally affected regime described by \citet{Ostilla2025}: the $\Gamma=1$ and $2$ cylinders show enhanced heat transport and reduced mean temperature, as in the periodic layer ($\overline{\langle wT\rangle} \approx 0.253$--$0.256$ and $\overline{\langle T\rangle} \approx 0.0241$ at $\Ek=3.16\times10^{-4}$), while the $\Gamma=1/2$ cylinder shows no such enhancement, since its non-rotating transport already matches the peak achieved in larger domains. With further rotation, $\overline{\langle wT\rangle}$ decreases and $\overline{\langle T\rangle}$ approaches the conductive limit in all configurations, and at $\Ek=3.16\times10^{-5}$ the cylinders remain close to the periodic layer.

The geometries diverge sharply in the wall mode regime, marked by the shaded band $\Ek \lesssim 2.3\times10^{-5}$ ($\mathit{Ro}^{-1} \gtrsim 4.3$). At $\Ek = 1.78\times10^{-5}$, the periodic layer has transitioned into a purely conductive state, whereas all cylinders maintain active convective heat transport ($0.05 \lesssim \overline{\langle wT\rangle} \lesssim 0.11$) and keep $\overline{\langle T\rangle}$ below conductive levels. Convection persists down to $\Ek = 5.62\times10^{-6}$—a factor of about four below the periodic layer's conductive boundary—with $\overline{\langle wT\rangle}$ decaying monotonically with decreasing $\Ek$ until all cylinders become quiescent at $\Ek = 3.16\times10^{-6}$. In this regime, $\overline{\langle wT\rangle}$ decreases systematically with increasing $\Gamma$, consistent with transport localised along the boundary, whose volumetric ratio scales as $1/\Gamma$. The sidewall mode thus sustains convection at rapid rotation rates where unbounded or laterally periodic domains are fully stabilised.

\begin{figure}
  \centering
  \includegraphics[width=0.8\textwidth]{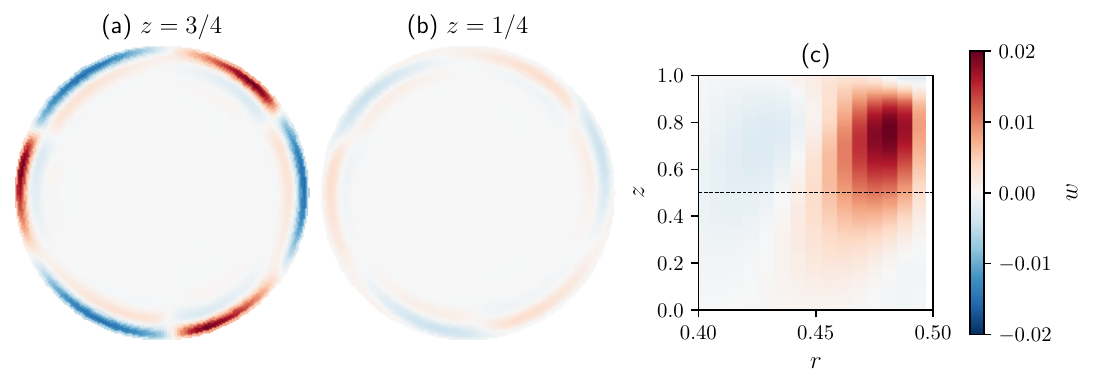}\\[2pt]
  \includegraphics[width=0.8\textwidth]{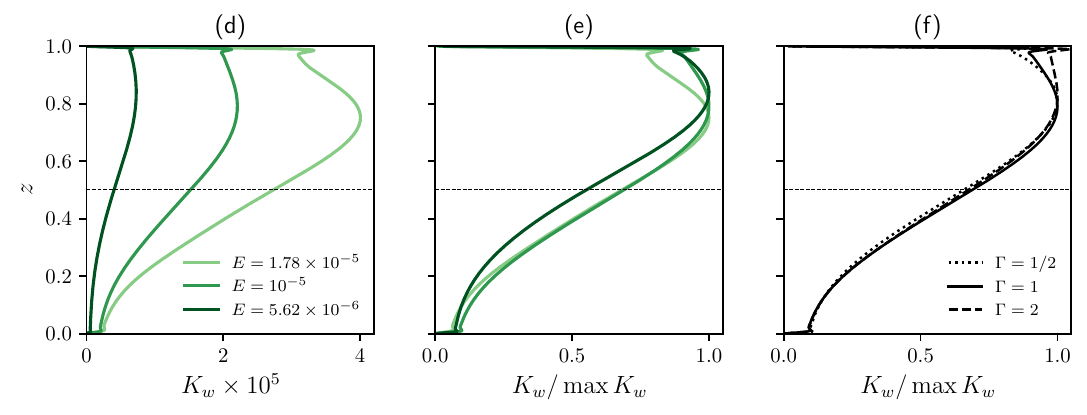}
  \caption{Vertical structure of the wall mode at $\Gamma=1$ and $\Ek=10^{-5}$ ($\Ra=10^8$). Vertical velocity $w$ shown in horizontal cross-sections at ($a$) $z=3/4$ and ($b$) $z=1/4$, and ($c$) in a radial--vertical plane sliced through the peak velocity lobe. Panels ($a$--$c$) share a common color scale. The mode exhibits azimuthal wavenumber $m=3$ and reaches maximum intensity in the unstably stratified upper half, while extending with reduced amplitude into the stably stratified lower half. ($d$) Fluctuation kinetic energy $K_w$ within the wall layer versus $z$ for $\Gamma=1$ at $\Ek=1.78\times10^{-5}$, $10^{-5}$, and $5.62\times10^{-6}$ (darker green denotes stronger rotation); the spikes near the plates are due to the Ekman layers. ($e$) The same profiles normalised by their maxima, and ($f$) normalised profiles for $\Gamma=1/2$, $1$, and $2$ at $\Ek=10^{-5}$.}
  \label{fig:vertical}
\end{figure}

We now examine the detailed spatial structure of the wall mode, focusing first on the cylinder with $\Gamma=1$. The mode is characterised by its dominant azimuthal wavenumber $m$, which is extracted from the azimuthal Fourier spectrum of vertical velocity $w$ at $z=3/4$ along the radius of maximum intensity, and by the fluctuation kinetic energy $K_w$ within the outer wall region ($r \ge 0.4$ for $\Gamma=1$), computed from volume normalised, azimuthally and time-averaged velocity variances. Figure~\ref{fig:vertical} illustrates this structure at $\Ek=10^{-5}$. In the horizontal plane $z=3/4$ (figure~\ref{fig:vertical}$a$), the vertical velocity field forms a perimeter ring of alternating lobes with $m=3$ attached to the sidewall, surrounding a quiescent interior core. At $z=1/4$ (figure~\ref{fig:vertical}$b$), plotted on the same color scale, the velocity lobes persist with reduced amplitude. A radial--vertical cross-section through a velocity peak (figure~\ref{fig:vertical}$c$) confirms that the mode forms a boundary attached layer that reaches maximum intensity in the upper half while extending well below mid-depth.

Vertical profiles of $K_w$ (figures~\ref{fig:vertical}$d$,$e$) quantify this distribution; the localized peaks near the top plate due to the Ekman layers are not considered. At $\Ek=10^{-5}$, the fluctuation kinetic energy peaks near $z=0.8$ and decays smoothly toward the lower plate; at mid-depth it retains roughly two-thirds of its maximum, and its average over the lower half is approximately $3.3$ times smaller than over the upper half. At $z=3/4$, the kinetic energy in the cylinder interior lies four orders of magnitude below its sidewall peak. The profile is almost identical at $\Ek=1.78\times10^{-5}$, while at the lowest Ekman number ($\Ek=5.62\times10^{-6}$) the amplitude diminishes and the energy peak shifts upward to $z=0.84$, with the mid-depth energy falling to half its maximum. Thus, while the mode is driven primarily in the unstably stratified upper half, it penetrates deeply into the stably stratified lower half. This vertical profile is independent of aspect ratio: at $\Ek=10^{-5}$, normalised profiles for $\Gamma=1/2$, $1$, and $2$ collapse onto a single curve (figure~\ref{fig:vertical}$f$). The active wall layer, defined where $K_w$ exceeds $10\%$ of its peak, has a constant radial thickness of roughly $0.03$ across all aspect ratios, confirming its sidewall nature. Only the absolute amplitude scales with geometry, increasing by roughly a factor of four from $\Gamma=1/2$ to $\Gamma=1$ before saturating at $\Gamma=2$.

\begin{figure}
  \centering
  \includegraphics[width=0.7\textwidth]{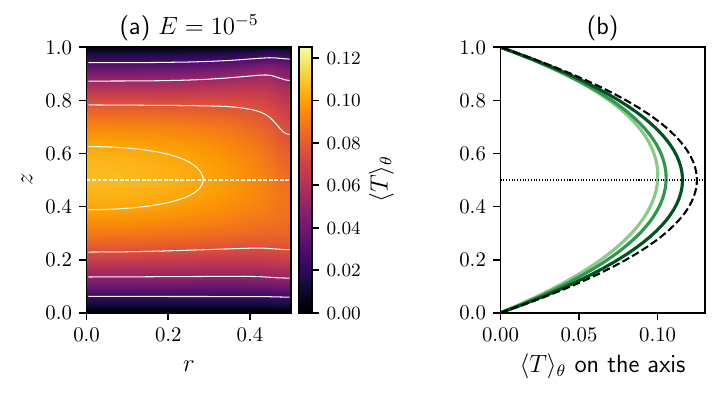}
  \caption{($a$) Azimuthally and time-averaged temperature $\langle T\rangle_\theta$ in the radial--vertical plane for $\Gamma=1$ at $\Ek=10^{-5}$ ($\Ra=10^8$); white contours represent isotherms at $0.025$, $0.05$, $0.075$, and $0.1$, and the dashed line indicates mid-depth $z=1/2$. ($b$) Vertical profile of $\langle T\rangle_\theta$ near the cylinder axis for $\Ek=1.78\times10^{-5}$, $10^{-5}$, and $5.62\times10^{-6}$, compared against the conductive profile $T_c(z) = \tfrac{1}{2}z(1-z)$ (dashed line); darker green corresponds to stronger rotation (lower $\Ek$).}
  \label{fig:Tmean}
\end{figure}

Analysis of the mean temperature field demonstrates that mode penetration into the lower half does not homogenise the fluid column. Figure~\ref{fig:Tmean} presents the azimuthally and time-averaged temperature $\langle T\rangle_\theta$ at $\Ek=10^{-5}$ for $\Gamma=1$. Away from the vertical sidewall, isotherms remain predominantly horizontal, and the centerline profile along the symmetry axis closely tracks the parabolic shape of the conductive state $T_c(z) = \tfrac{1}{2}z(1-z)$, albeit with a slightly reduced peak value. The convective heat transport generated by the wall mode thus cools the interior core slightly, though the bulk fluid approaches the purely conductive profile as rotation strengthens. Crucially, the lower layer remains stably stratified despite the deep penetration of the mode. Throughout the height interval $0.1 \le z \le 0.4$, the mean vertical thermal gradient $\partial_z \langle T\rangle_\theta$ remains positive at all radial locations. At $\Ek=10^{-5}$, this gradient retains approximately $84\%$ of its conductive value along the cylinder axis and $62\%$ near the sidewall, where mode amplitude is maximal. The sidewall mode therefore penetrates the stable lower region while retaining stable mean stratification in the measured lower-layer interval, producing only localised attenuation of the stable stratification near the boundary.

Finally, we analyse the azimuthal structure and drift velocity of the sidewall mode. For each configuration, approximately $200$ instantaneous snapshots of temperature in the horizontal plane $z=3/4$, separated by one free-fall time unit ($d/U_{ff}$ with $U_{ff}=\sqrt{\Ra\Pran}\,\kappa/d$, i.e.\ $10^{-4}\,d^2/\kappa$) and covering a total of about $0.02\,d^2/\kappa$, are evaluated. We subtract the azimuthal mean and compute the azimuthal Fourier decomposition in $\theta$ along the radius of maximum temperature fluctuation within the wall layer. The dominant wavenumber $m$ accounts for approximately $97\%$--$98\%$ of the fluctuation power across most cases (decreasing to $90\%$, $83\%$, and $75\%$ at $\Ek=5.62\times10^{-6}$ for $\Gamma=2$, $1$, and $1/2$, respectively). The drift frequency $\omega$ is determined from the phase evolution of this dominant Fourier component by finding the frequency that maximises the temporal coherence of the travelling wave pattern $\cos(m\theta - m\omega t)$. Here, $\omega$ is expressed in units of $\kappa/d^2$ in the frame co-rotating with the cylinder, with $\omega > 0$ denoting prograde drift. In all cases, the mode forms a coherent, rigidly rotating wave structure with temporal coherence exceeding $0.99$.

\begin{figure}
  \centering
  \includegraphics[width=0.8\textwidth]{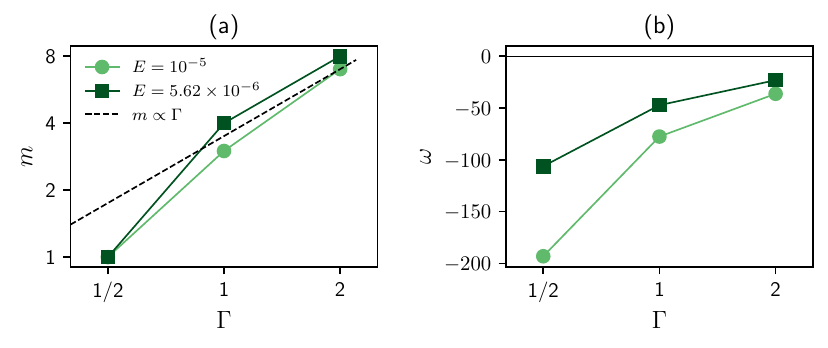}
  \caption{($a$) Azimuthal wavenumber $m$ of the sidewall mode and ($b$) its drift frequency $\omega$ (in units of $\kappa/d^2$ in the co-rotating reference frame; negative values denote retrograde drift) versus aspect ratio $\Gamma$ at $\Ra=10^8$ for $\Ek=10^{-5}$ and $\Ek=5.62\times10^{-6}$ (darker green denotes stronger rotation). The dashed line in ($a$) indicates $m \propto \Gamma$, corresponding to an aspect-ratio-independent wall wavelength. Both quantities are extracted from temperature fluctuations at $z=3/4$ along the radius of maximum intensity within the wall layer.}
  \label{fig:m_drift}
\end{figure}

Figure~\ref{fig:m_drift}($a$) shows that at $\Ek=10^{-5}$, the azimuthal wavenumber scales with aspect ratio, taking values $m=1$ at $\Gamma=1/2$, $m=3$ at $\Gamma=1$, and $m=7$ at $\Gamma=2$. The corresponding wall-parallel wavelength, $\lambda = \pi\Gamma d / m$, remains approximately $0.9$--$1.05\,d$ for $\Gamma \ge 1$ (and $1.6\,d$ for $\Gamma=1/2$, where $m=1$ represents the geometric lower bound). Consequently, $m$ grows roughly in proportion to the perimeter, i.e., to $\Gamma$. The mode drifts in the retrograde (anticyclonic) direction ($\omega < 0$) across all aspect ratios, drifting more slowly in wider cylinders (figure~\ref{fig:m_drift}$b$): $\omega = -1.9\times10^{2}$, $-78$, and $-36$ for $\Gamma=1/2$, $1$, and $2$. The passage frequency at a fixed wall location, $m|\omega|$, varies far less ($1.9\times10^{2}$, $2.3\times10^{2}$, and $2.5\times10^{2}$, respectively), indicating weak dependence on $\Gamma$. Reducing the Ekman number to $\Ek=5.62\times10^{-6}$ increases the wavenumber by one for $\Gamma=1$ and $2$ (to $m=4$ and $8$), shortening the wavelength to $\lambda\approx0.79\,d$ in both cylinders, whereas the narrow cylinder ($\Gamma=1/2$) retains $m=1$, with a larger fraction of the power shifting to $m=2$. The retrograde drift slows to $\omega=-1.1\times10^{2}$, $-47$, and $-23$ for $\Gamma=1/2$, $1$, and $2$, by factors of $1.8$, $1.6$, and $1.6$, close to the factor of $1.78$ by which $\Ek$ decreases, while $m|\omega|$ decreases to $1.9\times10^{2}$ at $\Gamma=1$ and $1.8\times10^{2}$ at $\Gamma=2$.

These kinematic properties share notable similarities with sidewall modes in rotating Rayleigh--B\'enard convection (RRBC). As in RRBC, the mode manifests as a boundary-attached wave drifting retrograde (anticyclonically) relative to the frame rotation \citep{Goldstein1993,deWit2023}. Near onset in RRBC, linear asymptotic theory predicts an azimuthal wavenumber $m \approx \left[\frac{\pi}{2}(2+\sqrt{3})^{1/2} - 17.49\,\Ek^{1/3}\right]\Gamma$ \citep{deWit2023}, which yields $m \approx 3\Gamma$ as $\Ek \to 0$, giving $m \approx 1.3$, $2.7$, and $5.3$ for $\Gamma=1/2$, $1$, and $2$ at $\Ek=10^{-5}$. Our observed values ($m=1$, $3$, and $7$) exhibit the same linear proportionality with $\Gamma$, albeit with a slightly steeper slope. Furthermore, the increase in $m$ with decreasing $\Ek$ observed at $\Gamma=1$ and $2$ is more pronounced than predicted by the linear RRBC expression ($m \approx 2.7$ and $5.4$ at $\Ek=5.62\times10^{-6}$ versus our numerical values of $m=4$ and $8$). Exact quantitative agreement is not expected, as the theoretical formula describes the linear onset threshold of standard RRBC, whereas our internally heated sidewall mode operates above onset and is driven exclusively in the upper half of the fluid layer. The radial thickness of the mode ($\approx 0.03\,d$ at $\Ek=10^{-5}$) remains comparable to the classic Stewartson boundary layer scales $\Ek^{1/3} \approx 0.02$ and $\Ek^{1/4} \approx 0.06$ that characterise sidewall structures in RRBC \citep{deWit2023}. The primary distinction lies in the vertical structure, which is non-uniform over the fluid height, and the underlying thermal driving mechanism. The drift is also of comparable magnitude. Normalised with the rotation rate, which is $\Omega=1/(2\Ek)$ in the present units, the angular drift velocities at $\Ek=10^{-5}$ are $\omega/\Omega\approx3.9\times10^{-3}$, $1.5\times10^{-3}$, and $7.3\times10^{-4}$ for $\Gamma=1/2$, $1$, and $2$, which is of the same order as the RRBC value at onset, $\omega_{dc}/\Omega\approx(4\pi^2[3(2+\sqrt{3})]^{1/2}\Ek-1465\,\Ek^{4/3})/\Pran\approx1.0\times10^{-3}$ \citep{deWit2023}; at $\Ek=5.62\times10^{-6}$ our values for $\Gamma=1/2$, $1$, and $2$ are $1.2\times10^{-3}$, $5.3\times10^{-4}$, and $2.6\times10^{-4}$, against $6.0\times10^{-4}$. A more quantitative comparison above onset would require defining an effective Rayleigh number for the internally heated system, which lies beyond the scope of the present study.

\section{Conclusions}\label{sec:conclusions}

Using direct numerical simulations, we have shown that rotating internally heated convection in cylinders supports a sidewall mode that sustains heat transport at rotation rates where a doubly periodic layer is conductive. At $\Ra=10^8$ and $\Pran=1$, the periodic layer is conductive at $\Ek=1.78\times10^{-5}$, whereas cylinders with $\Gamma=1/2$, $1$ and $2$ keep transporting heat down to $\Ek=5.62\times10^{-6}$, more than a factor of three lower, with a heat flux that decreases with $\Gamma$, consistent with transport localised along a lateral boundary. At weaker rotation the cylinders track the periodic layer, apart from a minor enhancement at $\Gamma=1/2$ that we attribute to domain confinement. The mode is a boundary-attached travelling wave with a thickness of about $0.03\,d$ independent of $\Gamma$ and a quiescent interior. Although the unstable driving is restricted to the upper half of the conductive state, the mode extends deep into the stably stratified lower half, with a vertical structure that is nearly invariant with Ekman number and aspect ratio; the lower half nevertheless keeps a positive mean temperature gradient, demonstrating that stable mean stratification is retained over the measured lower-layer interval.

Azimuthally, the wavenumber grows in proportion to $\Gamma$, and by one when $\Ek$ is lowered from $10^{-5}$ to $5.62\times10^{-6}$ at $\Gamma=1$ and $2$, which gives a wall wavelength of the order of $d$ independent of the cylinder size. The mode drifts retrograde as a rigid wave, more slowly in wider cylinders and at lower $\Ek$. These conclusions are based on a single $\Ra$ and $\Pran$, adiabatic sidewalls and three aspect ratios. Since the simulations start from the non-rotating state, they establish the existence and finite-amplitude persistence of the mode but not its linear onset threshold, which may differ if the instability is subcritical. The dependence on $\Gamma$, $\Ra$ and $\Pran$ of the wall mode is a topic for future linear stability analysis. Our findings nevertheless show that stability limits and heat transport inferred from laterally periodic domains must be re-evaluated when sidewalls are present, since the wall mode dominates the dynamics where the bulk remains stable.

\backsection[Acknowledgements]{We thank the Systems Unit of the Information Systems Area of the University of C\'adiz for computer resources and technical support.}

\backsection[Funding]{}

\backsection[Declaration of generative AI]{The authors used Claude Sonnet (Anthropic) to assist with the writing of the manuscript and with the production of the figures, including the analysis and plotting scripts. The authors reviewed and edited all AI-assisted content and take full responsibility for it.}

\backsection[Competing interests]{The authors declare none.}

\bibliographystyle{jfm}
\bibliography{references}

\end{document}